# NXNSAttack: Recursive DNS Inefficiencies and Vulnerabilities

Yehuda Afek[*][†]
*Tel Aviv University*
*afek@post.tau.ac.il*

Anat Bremler-Barr
*The Interdisciplinary Center*
*bremler@idc.ac.il*

Lior Shafir[†]
*Tel Aviv University*
*lior.shafir@gmail.com*

## Abstract

This paper exposes a new vulnerability and introduces a corresponding attack, the NoneXistent Name Server Attack (*NXNSAttack*), that disrupts and may paralyze the DNS system, making it difficult or impossible for Internet users to access websites, web e-mail, online video chats, or any other online resource. The NXNSAttack generates a storm of packets between DNS resolvers and DNS authoritative name servers. The storm is produced by the response of resolvers to unrestricted referral response messages of authoritative name servers. The attack is significantly more destructive than NXDomain attacks (e.g., the Mirai attack): i) It reaches an amplification factor of more than 1620x on the number of packets exchanged by the recursive resolver. ii) In addition to the negative cache, the attack also saturates the 'NS' section of the resolver caches. To mitigate the attack impact, we propose an enhancement to the recursive resolver algorithm, MaxFetch($k$), that prevents unnecessary proactive fetches. We implemented the MaxFetch(1) mitigation enhancement on a BIND resolver and tested it on real-world DNS query datasets. Our results show that MaxFetch(1) degrades neither the recursive resolver throughput nor its latency. Following the discovery of the attack, a responsible disclosure procedure was carried out, and several DNS vendors and public providers have issued a CVE and patched their systems.

## 1 Introduction

The Domain Name System (DNS) infrastructure, a most critical highly dynamic system on which almost any access to a resource on the Internet depends, has recently been an attractive target for a variety of DDoS attacks [4, 31]. As seen in the Mirai attack [4], a degradation or outage of part of the DNS service disrupts many popular websites such as Twitter, Reddit, Netflix, and many others, impacting millions of Internet users. Moreover, recent large scale attacks, known as NXDomain attacks [33] (or *water-torture* attacks [20, 31]), directly tried to take down parts of the DNS system by flooding the DNS servers with well-structured requests of pseudo-randomly generated nonexistent sub-domains (PRSD).

This paper explores the inefficiencies and vulnerabilities of recursive resolvers. We analyze the DNS recursive resolver behavior and the interaction between its algorithms and components using the popular BIND [15] server implementation. We expose a new vulnerability in recursive resolver algorithms and demonstrate a new attack, called *NXNSAttack*, which exploits this vulnerability. Finally, we suggest and analyze modifications to the recursive resolver algorithms, called MaxFetch(1) and Max-Breadth, which drastically reduce the effectiveness of this attack.

At an abstract level, the DNS system has two parts, each of which is a large, highly distributed system: a hierarchical and dynamic database of *authoritative* name servers storing the DNS data, and a large number of client-facing *resolvers*, located either locally at the service providers and local organizations, or as cloud public services (e.g., CloudFlare 1.1.1.1, and Google 8.8.8.8) that walk through the hierarchical structure to retrieve the domain name resolutions to IP addresses. The focus of the current paper is on the interaction between the recursive resolvers and the authoritative hierarchical structure.

In walking through the authoritative hierarchy, the resolver is delegated from one authoritative server to another. The delegation messages are called name server (NS) referral responses. In such a referral message, an authoritative server tells the recursive resolver that it does not know the answer to its query and refers it to another name server. One of our main observations is that the information in the NS referral responses, at the different recursive steps, and the actions taken by the recursive resolvers as a result, may introduce huge communication and other resource overheads.

[*]Member of the Checkpoint Institute of Information Security. Partial support provided by the Blavatnik Interdisciplinary Cyber Research Center, (ICRC).
[†]Supported by the Blavatnik Family Grant.

These overheads occur mainly because the name servers in the NS referral response are not always provided with their corresponding IP addresses (known as glue records). Top-level authoritative domains (TLDs), second-level domains (SLDs), and other authoritative servers are not allowed to provide IP addresses for domains that do not reside in the same zone origin (known as *Out-of-Bailiwick* name servers [14]). This is mostly to protect from DNS poisoning attacks.

We study the implications and prevalence of this phenomenon. We first discuss (§2) our observation that the number of packets involved in a typical resolution process is much larger in practice than expected in theory, mainly due to proactive extra resolutions of name server IP Addresses.

We then show how the proactive resolution of *all* the name servers in the referral response becomes a major bottleneck in recursive servers such as BIND, considered as the de facto standard for DNS software. We present a new attack, called *NXNSAttack* (§3), that exploits this vulnerability and is more effective against authoritative and recursive servers than the NXDomain attack (§4.5). We show three variants of this attack (*a*, *b*, and *c* in Table 1), analyzing their impact on a BIND based recursive resolver and authoritative servers. (§4). The NXNSAttack simulations saturate the recursive resolver's cache (with NX & NS records) and reach a packet amplification factor (PAF) of more than 1600x (variant *a*). The key enabler for the attack is the ease with which an attacker can acquire and control an authoritative server.

|  |  | Attack target (victim) | Max Amplification factor | |
|---|---|---|---|---|
|  |  |  | Bytes | Packets |
| NXDomain Attack (Mirai [4]) |  | Authoritative name server | 2.6x | 2x |
| NXNSAttack | a | Recursive resolver | 163x | 1621x |
|  | b | Authoritative SLD | 21x | 75x |
|  | c | Root / TLD | 99x | 1071x |

Table 1: Three variants of the NXNSAttack, and the NXDomain attack [33] empirical evaluation under BIND 9.12.3

We then show how the BIND DNS resolver algorithm can be enhanced to remove unnecessary proactive fetches (§5), thus alleviating the vulnerability, and measure the performance improvements. In particular, we show that our MaxFetch(1) enhancement has no negative impact on either the latency or throughput of the enhanced recursive resolver.

Finally (§6), we quantify the pervasiveness of domains with *out-of-bailiwick* name servers in: (i) the top million domain resolutions, and (ii) in a campus DNS traffic trace. Since the inefficiencies and vulnerabilities we uncover are associated mostly with referral responses that contain many name servers without an associated IP address, we study the prevalence of the phenomenon. We find that in 60% of the domains, all the name servers are *out-of-bailiwick* .

Related work is discussed in §7, responsible disclosure is reported in §8, and our conclusions are given in §9.

## 2 Background: DNS Resolution Process Overhead

Continuing the description given in the introduction, the main concern in this paper is the interaction between a resolver (of which there are millions in the Internet) and the authoritative name servers (of which there are more than 10 million) in the process of retrieving the required resolution from the authoritative servers. These authoritative servers are authorized to provide the DNS data (translating domain names to IP addresses) for a specific zone without performing requests to other DNS servers.

Cache memory at the resolver side plays a critical role in significantly reducing the amount of interaction between the resolvers and the authoritative hierarchy. By recording previous resolutions for a period of time rather than querying an authoritative server again, the information is retrieved from the cache. However, the vulnerability and attacks we discover in this paper bypass the cache by making sure to query about domain names that are not present in the cache. Therefore we analyze the system behavior with an empty cache. Cache records and DNS response records are tagged by either one of the following labels: A, AAAA, NS, or NX indicating the type of information they carry: IPv4 address of a particular domain, IPv6 address, authoritative name-servers for a domain or zone, a domain name that does not exist in the appropriate authoritative server, respectively.

### 2.1 The Resolution Process: In Theory

In a clean and fault-free system, when an answer is not found in the resolver cache, it walks through the authoritative hierarchy to obtain it, as shown in Figure 1, where a recursive resolver resolves the domain name `www.microsoft.com`. It starts by issuing a query to one of the root servers (e.g., A.ROOT-SERVERS.NET, whose IP address is hard coded into the recursive resolver), asking for the address of `www.microsoft.com` (step 1 in Figure 1). The root server returns an NS referral response delegating the query to one of a few TLD (Top Level Domain) name servers responsible for the '.com' zone (step 2). The recursive resolver selects one of these name servers and issues another query (step 3) asking the chosen TLD name server (assuming it has its IP address) for the address of `www.microsoft.com`. The .com TLD server responds with another NS referral response (step 4) delegating the query to one of a few SLD (Secondary Level Domain) name servers responsible for the 'microsoft.com' zone. The recursive resolver again selects one of these name servers and issues another query asking for the address of `www.microsoft.com` (step 5). The SLD authoritative server owns the DNS records for all the domains under 'microsoft.com'; and returns an 'A' response with the requested IP address (step 6). Thus, after 3 rounds of query-response between the resolver and the authoritative servers, the final answer is obtained and is forwarded to the querying client.

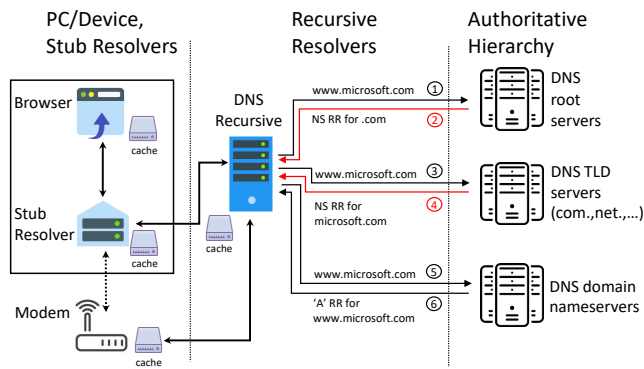

Figure 1: The resolution process, in theory, for the domain www.microsoft.com. The red steps represent NS referral responses.

## 2.2 The Resolution Process: In Practice

Here we show that in practice the resolution process requires many more messages to be exchanged between the resolver and the authoritative servers due to fault-tolerance and low latency requirements. We analyze hundreds of thousands of resolutions taken from top websites and campus DNS data, inspecting the type and number of packets involved in each resolution. We tested a BIND 9.12.3 recursive resolver installed on an AWS EC2 machine, as well as on a local machine, to inspect the code and to analyze its internal components and algorithms. We discovered that while the procedure described in Figure 1 results in a total of three requests and replies, in practice it results in many more messages (see the procedure described in Figure 2), sometimes hundreds, even if the cache has been filled by many previous but different requests.

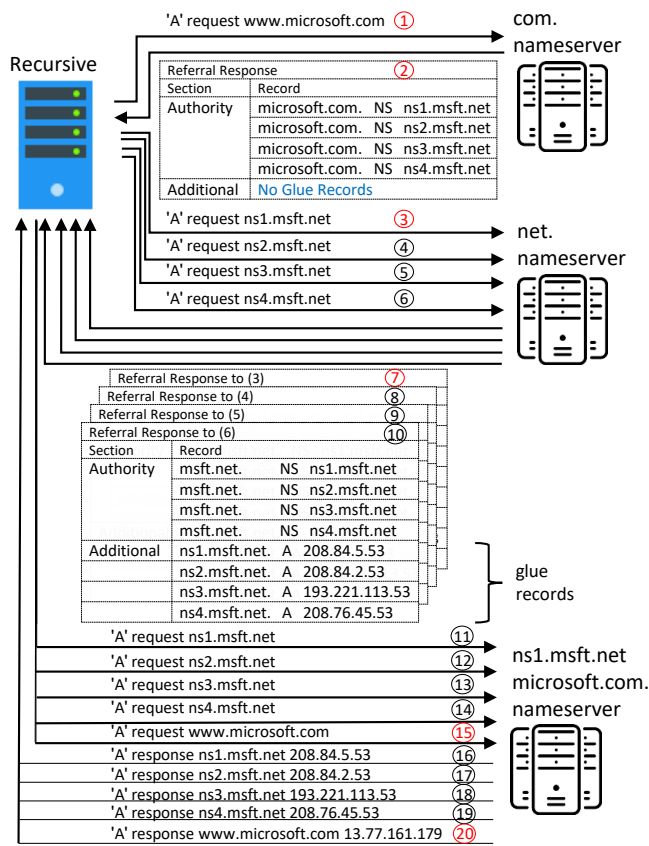

Figure 2: The resolution process in practice, for the domain www.microsoft.com by the BIND 9.12.3 recursive resolver (parallels the diagram in Fig. 1). The .net and .com TLD name servers are already cached at the beginning of the process. The red steps are the mandatory messages required to answer the client query.

For example, `microsoft.com` resolution requires 54 IPv4 packets (actually 126, but we exclude TCP handshake and control packets that are used when the responses are too large due to additional records and EDNS), `twitter.com` resolution requires 388 packets and `www.gov.uk` requires 102. A detailed analysis of two data sets carried out in §5.3 shows for example that 61.56% of the 'A' queries for the top million domains result in considerably more messages than expected in theory.

**Name server referral response:** In resolving a domain name, when the cache is empty, the resolver traverses the authoritative hierarchy starting from the root server. It is delegated from one authoritative server to another, until reaching the authoritative server that has the required mapping of the domain name in question to an IP address. The delegation messages are called name server (NS) referral responses. In such a referral message, an authoritative server tells the recursive resolver: "I do not have the answer, go and ask one of these name servers, e.g., ns1, ns2, etc., that should get you closer to the answer".

To provide fault tolerance, the information at each level of the hierarchy must be duplicated on several authoritative name servers. The root has 13 copies (each duplicated using anycast, to a total of more than one thousand actual servers). All authoritative servers must have at least 2 copies, and 99% have less than 7 copies; see Figure 14.

The referral response message provides the copies of the authoritative server that the resolver is delegated to by their domain names (see examples in Figures 2 and 4). In addition, sometimes the referral response also provides the IP address of each copy, called the *glue record* of the corresponding authoritative name server. These glue records are provided in 'A' records within an NS record, in the referral response, and may be present for none, some or all the name servers in an NS record. The DNS specifications do not provide clear guidelines on when glue records should be present nor how to process them on the recursive side. By RFC 1034 [24] glue records are required only if the NS is lying within or below the zone or domain for which it acts as a name server. Consider for example, 'ns5.google.com' in zone 'google.com'. This condition is known as the *Bailiwick* rule, or more specifically *in-bailiwick'*. This requirement was introduced to avoid a query deadlock for NS referrals that contain name servers within the domain being queried. For example, if the recursive resolves `www.example.com` and the TLD returns a referral containing `ns5.example.com` as a delegated name server which resides within the example.com domain, but without its IP address, the recursive will then initiate another A query asking to resolve `ns5.example.com`. It will be again referred to `ns5.example.com`, which leads to a live-lock (infinite loop).

Another important motivation for the *Bailiwick* rule is to avoid and reduce the risk posed by cache poisoning attacks [32, 34]. In such attacks, the owner of any DNS authoritative server could hijack ownership on any domain name. When responding to a query from a recursive resolver, such a malicious authoritative server can send an NS referral record that includes any domain name as a NS with a glue record mapping this domain name to any IP address, thus injecting or overriding a bogus A record for any domain into the recursive resolver cache. To prevent such cache poisoning attacks using malicious glue records, many recursive implementations store glue records as 'A' records in their cache only if they comply with the *Bailiwick* rule. Otherwise, in an *out-of-bailiwick* case, for example, ns.example.net as a name server for the example.com zone, the resolver discards the glue record. Generally, without getting into different variations and implementation details, the BIND [15] recursive implementation, which we analyze in this paper, as well as Unbound [19], PowerDNS [2] and Microsoft DNS, all discard out-of-bailiwick glue records. Other solutions to eliminate cache poisoning attacks as a result of out-of-bailiwick glue records include DNSSEC, which authenticates the authoritative responses by verifying their signature through a chain of authority. However, these have a very low adoption rate.

Another important consideration that influences the cost of a resolution with an empty cache is the requirement to minimize the response time. The resolver attempts to resolve the domain name of each name server in the referral response for which it does not have an IP address, immediately upon receiving the referral message. Thus, if for example the referral response delegates the recursive to one of 30 name servers for which it does not have an IP address, the recursive immediately starts (BIND implementation) 30 new resolutions. This ensures that the resolver has the IP address of each authoritative it may need, as soon as possible, without incurring additional delays. In addition, most of the recursive resolver implementations use algorithms to distribute the load among the different name servers and achieve lower latency over time when sending queries to authoritative name servers. For example, BIND uses an sRTT (smoothed Round Trip Time) algorithm with a decaying factor, in which it tracks the response time of each name server. Other algorithms perform an initial round-robin over the name servers followed by measured latency-aware selections.

Figure 2 illustrates the additional out-of-bailiwick requests that the recursive issues for `www.microsoft.com`. In this case, the TLD (.com and .net) name servers are already in the cache as a result of previous requests. The .com authoritative responds with an NS referral (step 2) containing four out-of-bailiwick name servers (`ns*.msft.net` for the queried zone microsoft.com). The recursive then initiates four additional resolution fetches for all these out-of-bailiwick name servers. Note that even after it receives their IP addresses in the referral responses (steps 7-10) as glue records, it still performs additional resolution requests for them (steps 11-14). This is because their corresponding requests' recursion state was already initiated independently with an indication that they are not cached.

Note that we observe additional causes for the high number of messages in DNS resolutions: (i) too long NS responses that include multiple name servers and other options such as RRSIG and NSEC3 data in the additional records, leaving no place for all the glue records in a 512-byte UDP packet, forcing the recursive to resend the request using TCP, or by using the UDP EDNS0 4096-byte option. (ii) Canonical NAME records (CNAME) that reside in different domains than the queried one, and which sometimes have to be resolved with an additional fetch starting from the root-servers.

In conclusion, the referral procedure results in proactive additional resolutions for *all* the non-cached name servers that appear in the NS referral response that are either out-of-bailiwick or do not have an associated glue record. This rule is not part of the configuration nor can it be disabled. In this paper, we focus on these extra resolutions and propose a change in the way they are handled. We claim that the resolution of the referred name servers should be distributed and amortized over several client requests (see §5), in contrast to the current practice where all the resolutions are performed in the first client request. Moreover, many domains outsource their authoritative name servers to cloud operators such as Cloudflare, Google.com, or domaincontrol.com, and these operators often choose short TTL values (30 or 60 seconds). This in turn causes many server resolutions to be outdated by the time the resolver wants to use them. As a result, the resolver has to redo the corresponding resolution(s).

The gap between the number of resolution packets per query expected in theory and the number observed in practice raises several issues that we address in the following sections:

1. In §3 we expose a new vulnerability and corresponding attack, the NXNSAttack.

2. In §5 a solution to mitigate the NXNSAttack by not resolving all the missing name server IP addresses in the first client query is suggested. The extra queries are amortized over future client queries.

3. In §5 we evaluate our solution, and present our experiments and measurements.

4. In §6 we measure the prevalence of *Out-of-Bailiwick* domains, and measure the overhead of additional *Out-of-Bailiwick* resolutions on two different datasets: (i) the top million domains list; (ii) a campus DNS trace.

## 3 NXNSAttack

Here we show how the multi-name server referral response and the resulting extra resolutions may be used to mount a new attack, NXNSAttack, on different elements of the DNS infrastructure.

As shown in the previous section, for each name server name without an associated IP address, in the NS referral response, the recursive resolver initiates a new resolution procedure. This is the core of our attack. The attacker uses an authoritative server that it owns to craft a response to a resolver with a referral that contains $n$ new and nonexistent name server names without an associated IP address, thereby causing the resolver to start the process of $F$ new resolutions. As shown later, the maximum $F$ can be in the range, $74 \leq F \leq 2 \cdot n$, where $n$ is the number of name server names in the referral response (in the BIND implementation, $2n$ requests to resolve the IPv4 and IPv6 address of each). When the attacker generates many such referral responses repeatedly, this results in a DDoS attack on either the resolver or on a corresponding authoritative server, with an amplification factor of $O(F)$ packets, sometimes much larger than $F$. There are several parameters and variants of this basic principle that we investigate in this paper.

### 3.1 Threat Model

To mount a NXNSAttack on either a recursive resolver or an authoritative server, an attacker should:

1. Have access to one or more DNS clients on the Internet. The attacker may use a botnet, like the Mirai IoT botnet [4], or an ad network [17].

2. Own or compromise an authoritative name server. An adversary who acts as an authoritative server has the ability to craft any NS referral response as an answer to different DNS queries. It controls the information that appears in the referral response, such as the number of name servers, their names, and their glue records (as well as the absence of glue records).

Controlling and acquiring a huge number of clients and a large number of authoritative NSs is not difficult. Authoritative name servers are easily and cheaply acquired by first buying and registering new domain names (for our experiments, we purchased several domain names for less than $1 each in less than 5 minutes). These acquired domain names can be dynamically associated with any authoritative server in the Internet. Alternatively, attackers today are able to compromise DNS operators' credentials and manipulate zone-files, sometimes even gaining access to their registrar records, as exemplified by recent DNS hijacking attacks [10, 23]. It should be noted that recent attacks have also utilized capabilities that are much harder to acquire, such as [4, 31] IoT botnets, Booters (DDoS for hire services [16]) and dynamic C&C servers.

### 3.2 The Amplifier

The core building block of the NXNSAttack is the *amplifier* (Figure 3), which is composed of two attacker components and one innocent recursive resolver. The two attacker components are a client and an authoritative name server. The attacker issues many requests for sub-domains of domains authorized by its own authoritative server (step 1). Each such request is crafted to have a different sub-domain in order to bypass the resolver's cache, thus forcing the resolver to query the attacker's authoritative server in order to resolve each of these sub-domains (step 2). The authoritative name server then returns an NS referral response with $n$ name server names without their glue records (step 3), i.e., without their associated IP addresses. This forces the resolver to start a resolution query for each one of the name server names in the response, regardless of whether they are *in-bailiwick* or *out-of-bailiwick*, because it does not have their IP addresses in its cache (step 4). The attacker's authoritative referral response issues $n$ new and different delegated name server names each time it receives a query for a sub-domain from the recursive resolver.

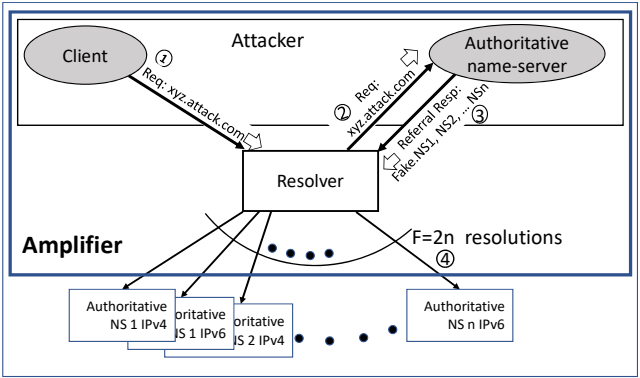

Figure 3: The amplifier: components and operation steps.

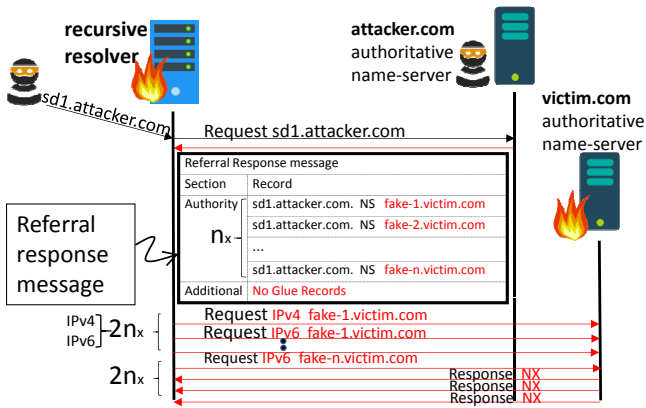

Figure 4: A different view of the messages exchanged in the steps of the amplifier operation in Figure 3.

The attacker can use the amplifier to generate different attacks on different targets by combining the following

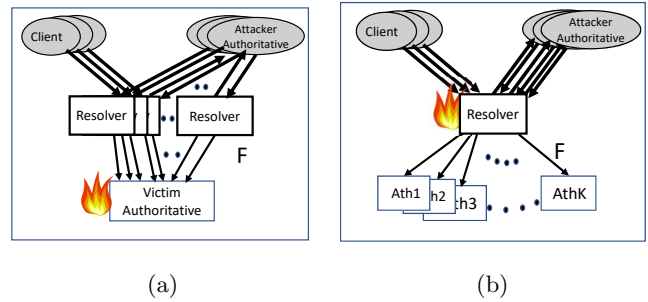

Figure 5: NXNSAttack targeting the authoritative server (a) and the recursive resolver (b)

three ingredients in a variety of ways:

**The bogus name server domains** placed by the attacker authoritative in the referral response determine the target authoritative server, which could be at different levels of the DNS hierarchy.

**Multiplicity of clients and/or resolvers** to target a single authoritative (Fig. 5a), or **multiplicity of authoritatives** to target a particular recursive resolver (Fig. 5b).

**Self delegation** by the attacker that places $n_1$ self-delegations to fake name servers in its own domain, in the first malicious referral response. The resolver then sends $F1 = 2n_1$ new requests to the attacker authoritative, which then crafts and issues $F1$ new referral responses (see Fig. 6), each of which contains $n_2$ delegations of the attacker's choosing. This results in a total of $2n_2 \cdot F1$ name server resolution requests issued by the resolver, thus doubling the attack fan out.

Here we focus on three basic attacks: against a recursive resolver, against an authoritative SLD victim (e.g., victim.com name server), and against the ROOT/TLD servers (.com, and "." ). See Table 1 for a summary of the amplification factors.

**Recursive resolver attack.** (Fig. 5b) Here the maximum packet amplification factor (PAF) is 1620x (both according to our model and empirically; see §4), achieved when the referral delegations are to different TLD name servers (e.g., `fake1.com`, `fake2.com`, ...,`fake1.net`, ...). For each two packets – one from the client and one from the authoritative name server – that the attacker components generate, the victim recursive resolver processes 3,242 packets, out of which 1,081 are DNS packets and the rest are TCP connection control packets. The corresponding bandwidth amplification (BAF) is 132x; see §4.3.

**Authoritative SLD attack.** (Fig. 5a) In this attack, all the name servers in the malicious referral are sub-domains of a victim SLD (second-level domain, e.g., `fake-1.victim.com`, `fake-2.victim.com`, ...). The maximum packet amplification factor is 74x, and the corresponding bandwidth amplification factor is 21x; see §4.3 for the cost and amplification factor analysis.

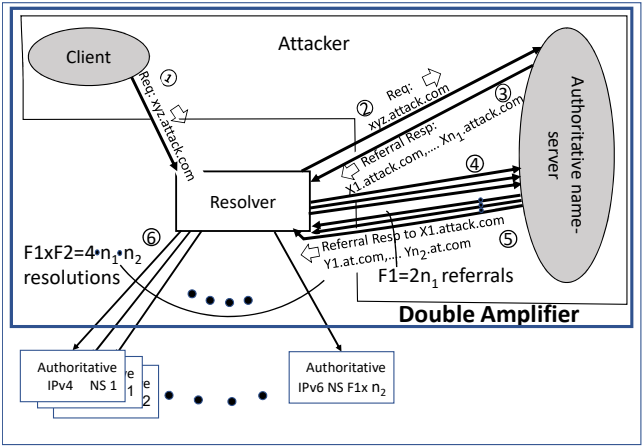

Figure 6: Illustration of the double amplification attack using self delegations in the first referral response. This attack variant (c) reaches a firepower of $F = F_1 \cdot F_2 = 37 \cdot 2 \cdot 135 \cdot 2 = 19,980$ (see §4.1).

**ROOT/TLD attack:** Here the attacker uses the self-delegations technique (Fig. 6) to increase the number of concurrent referrals to the ROOT name servers. In our empirical tests, the victim processes up to 81,428 packets (14,126,945 bytes) for each client request (and corresponding 75 referral packets) that the attacker generates (it is "only" 81,428 because many were lost). The high victim cost is because the first referral response from the attacker contains delegations to 37 new and different sub-domains of the attacker (e.g., `sd1.attacker.com`, ..., `sd37.attacker.com`), which results in 74 more requests (IPv4 and IPv6 for each delegated ns) to the attacker from the recursive resolver. The attacker's authoritative name server then responds with 74 crafted referrals, each containing 135 delegations to the ROOT server (e.g., `domain.fake` or `domain.tld`, where the TLD name servers are not cached in the recursive resolver), which in turn receives 18,980 concurrent requests; see §4.3 for the cost and amplification factor analysis.

## 4 NXNSAttack Analysis Evaluation

### 4.1 $F$, the Amplifier Firepower

The traffic fan-out of the amplifier as a result of one client request is measured by either the number of generated resolution requests, or the number of packets sent, or the number of bytes (bandwidth, bw) sent. In this section we present the corresponding numbers that were measured in our testbed setup and an analysis that explains them.

Recall that if the recursive resolver receives a referral response that delegates the original request to $n$ name server names, without providing their IP address (no glue records), it then generates – in theory – $2n$ requests to resolve IPv4 and IPv6, for each of these $n$ names. However, two parameters limit this number. The first is the maximum number of delegation names that fit into the referral response, denoted $n_{max}$ or just $n$. Our experiments show that $n_{max}$ is a function of the DNS packet size (including EDNS(0) extensions [6] and DNS over TCP) and the number of characters in the domain names. In our tests $n_{max}$ turned out to be 135. The second parameter is the *max-recursion-queries* parameter that, in Bind, sets the maximum number of requests a recursive resolver can send when resolving one client request. As stated in the BIND 9.12 manual [1]: "max-recursion-queries: Sets the maximum number of iterative queries that may be sent while servicing a recursive query. If more queries are sent, the recursive query is terminated and returns SERVFAIL. Queries to look up top level domains such as '.com' and '.net' and the DNS root zone are exempt from this limitation. The default is 75". We denote max-recursion-queries as $Max\_rq$.

Since in step 3 in Fig. 3 the recursive sends one request, the remaining $Max\_rq$ budget is 74. This is sufficient to resolve 37 names, requesting separately the IPv4 and IPv6 address of each, resulting in $F = 74$, unless the requests are sent to either the root or a TLD name server, in which case, $n_{max}$ is the only limiting factor, yielding $F = 2n_{max} = 270$.

### 4.2 Experimental Setup

We deployed an experimental setup like the one shown in Figure 3, on the AWS cloud in Ohio. Note that testing which involves attacking live operational name servers is not permissible. The setup includes a client, a recursive resolver, and two authoritative servers: one for the attacker and one for the victim. For each component, we used a large EC2 machine with 16Gb RAM and 4 vCPUs. The authoritative and recursive resolver servers run BIND 9.12.3 in authoritative and recursion operation mode respectively. The client is deployed on a different machine, configured to send DNS requests directly to our recursive resolver.

We chose BIND because it is the most prevalent DNS server implementation [15, 25] and is considered as the de facto standard for DNS servers. Moreover, a recent work [18] shows that the majority of open DNS resolvers operate BIND. We tested multiple versions of BIND in our experiments (different minor versions of 9.11 and 9.12), with no notable differences.

To show that the vulnerability is not unique to BIND, we also provide in §4.4 our results on open recursive resolvers including Google, CloudFlare, Dyn and others. All the open resolvers that we tested exhibited considerable amplification when sending a single NXNSAttack request.

## 4.3 Cost and Amplification Analysis

In Subsection 4.1 we computed $F$, the amplifier firepower, which is the total number of DNS requests generated by the amplifier, which was $2(\min(n,(Max\_rq-1)/2))$ if the attack is on an SLD domain, and $2n$ if the attack is on a TLD or on root servers (results in $F = 74$ and $270$ respectively). The 2 factor here is due to requesting the IPv4 address and IPv6 separately. But how many packets and bytes does it translate into? We measure it in our setup and explain (calculate) the observed numbers by analyzing the BIND protocol.

We claim that the cost to the victim in packets, denoted $C_v^{pkt}$, as a result of one client request, as a function of $F$, is:

$$C_v^{pkt} = 2 \cdot F \cdot (1 + 5 \cdot TC), \qquad (1)$$

where $TC$, the value of the truncate bit in the DNS protocol, equals 1 if the $F$ requests fall back to TCP, and 0 otherwise. The $TC$ bit indicates whether the UDP DNS request/response failed due to UDP packet size limitation and is retried in TCP. This often happens when the delegated name servers support DNSSEC signing (e.g., TLD servers, as we observed in our evaluation in §4.3). In such cases, the resolver retry (request and response) involves additional TCP control packets. In our evaluation in §4.3 we observe that each such request response exchange over TCP involves a total of 10 packets: DNS request, DNS response, and 8 TCP control packets (3 for handshake, and 5 for session termination).

The 2 factor in (1) is because we count both the packets sent and received by the recursive resolver or the authoritative victim towards their attack-cost. Traditionally, in DDoS bandwidth attacks, the packets/bytes amplification factor is the number of packets/bytes that are sent to the victim divided by the number of packets/bytes the attacker has sent. The victim name server is forced to receive many packets, process them, access memory, consume cache/memory capacity, and respond with a new DNS request or response packets including TCP connections. Therefore, our analysis of the amplification factor considers the packets the victim (the recursive or the authoritative) receives and sends.

Equation (1) provides $C_v^{pkt}$, the cost incurred by the victim (recursive resolver or authoritative server) when attacked by the amplifier. The packet amplification factor (PAF) of the different attacks is calculated by dividing the victim cost by that incurred by the attacker, denoted $C_a^{pkt}$. In attacks a and b (following Fig. 3) the attacker sends two messages, the client request and the referral response from the attacker-controlled authoritative name server. In attack c, Fig. 6, the attacker's authoritative server sends 74 packets, yielding $C_a^{pkt} = 75$.

PAF is the ratio between the number of packets processed by the victim and the number of packets sent by the attacker, i.e., $PAF = \frac{C_v^{pkt}}{C_a^{pkt}}$. Similarly, the bandwidth amplification factor is $BAF = \frac{C_v^{bw}}{C_a^{bw}}$, where $C_a^{bw}$ denotes the number of bytes that the attacker must send and $C_v^{bw}$ denotes the number of packets that the victim must process.

The costs discussed above are incurred with every client request because the attacker's authoritative server issues referral requests with new and different fake (nonexistent) names each time. In addition, there are one-time costs that we ignore but will show up in our measurements. These represent the packets exchanged between the recursive resolver and the ROOT/TLD authoritative name servers in resolving the attacker and the victim name servers, respectively. Since these name servers are cached after the first client request, we do not consider them in the packet cost analysis.

In Table 2 and below we analyze each attack variant, describing it and comparing its measured cost to its calculated cost according to the model above.

**(a) Recursive resolver attack**  (row a in Table 2). Here each attacker's referral (step 3 in Figure 3; see also Figure 4) contains delegations to many new and different name servers of the `.com` zone. The zone file contains millions of NS records and looks like this:

```
ORIGIN sd0.attacker.com.
sd0.attacker.com. IN NS ns1.fakens0.com.
sd0.attacker.com. IN NS ns1.fakens1.com.
...
sd0.attacker.com. IN NS ns1.fakens-n.com.
```

Considering that .com and other TLD name servers are external to our setup, we initiated only a few requests for `sd*.attacker.com`, while monitoring the recursive resolver behavior.

In our setup, $n_{max}$ turned out to be 135. The resulting firepower is thus 270 requests that are sent to one of the `.com` TLD name servers, asking 'who is `ns1.fakens*.com`?'. The `.com` name server responds with negative responses (NXDOMAIN). However, all TLD responses also contain a SOA record, RRSIG and multiple NSEC3 signatures (DNSSEC signatures), thus exceeding the maximum response size of 512 bytes. As a result, the TC bit is set to on, forcing the resolver to repeat the 270 queries over TCP (which also creates a lot of overhead on the resolver and the authoritatives to handle these TCP connections). Thus, $C_v^{pkt}$ by equation (1) is 3240. However, in our setup we measured 3243 due to the initial one-time resolution of the attacker's authoritative server and the victim recursive resolver addresses. The PAF is thus $\frac{C_v^{pkt}}{2} = 1620$.

| | Victims | Cost Factors | | Packets Cost | | PAF | Bytes Cost | | BAF |
|---|---|---|---|---|---|---|---|---|---|
| | | Firepower ($F$) | TC bit retry TCP | Attacker $C_a^{pkt}$ | Victim $C_v^{pkt}$ | | Attacker $C_a^{bw}$ | Victim $C_v^{bw}$ | |
| a | recursive resolver, TLD name server | 270 | 1 | 2 | C 3,240 M 3,243 | **1620x** | 3,967 | 647,107 | **163x** |
| b | SLD name sever, e.g., victim.com | 74 | 0 | 2 | C 148 M 150 | **74x** | 1,049 | 22,073 | **21x** |
| c | ROOT or TLD NS | 19,980 =74x270 | 1 | 76 | C 239,760 M 81,428 | **3154x** 1071x | M 142,487 | M 14,126,945 | M **99x** |

Table 2: Cost of different attack variants as a result of one client request, using BIND (M, measured cost; C, calculated cost).

The BAF in our setup was measured to be 163, very close to its expected value when taking into account the sizes of the different packets.

Note that here the .com TLD can also be considered as a victim because it processes the same packets as the recursive resolver under attack. Moreover, as described in Figure 5a, several resolvers may be used to mount a massive attack on any TLD or root server. We also performed this experiment with other TLDs (.live and .online) and received the same results.

**(b) Authoritative SLD attack:** To attack a particular SLD server, each attacker's referral is crafted to contain delegations to many new and different sub-domains of the victim SLD (e.g., fakens1.victim.com, fakens2.victim.com, ...).

In this attack variant, the BIND *max-recursion-queries* threshold does limit the number of iterative requests to 75. To test this attack we used two name servers, one as the attacker's and one as the victim. Since our authoritative victim does not use DNSSEC, no TCP retries are involved. Thus, $C_v^{pkt} = 2 \cdot 74 = 148$ and PAF is $\frac{C_v^{pkt}}{2} = 74$x. The victim bytes cost is $C_v^{bw} = 22,073$ bytes, and the attacker cost is 1,049 bytes, which leads to a BAF of 21x. As before, the measurement on one client request is 150 rather than 148 due to the one-time resolution of the attacker and resolver servers, which should not be counted towards the PAF or BAF calculations. Note that, had the authoritative victim used DNSSEC, the packet cost would likely have increased 6-fold according to equation 1, to 888, and the PAF to 444.

**(c) ROOT TLD attack.** To attack a TLD or root servers (a tough challenge since there are hundreds of them), one can try variant a, or try this variant with many fewer client requests, as described; see §3.2 and Fig. 6. Here the attacker uses the self-delegations technique to double the amplification factor in attacks against the ROOT or TLD name servers, in which the resolver is also victimized. The attacker's first referral (step 3 in the figure) contains $n_1$ different sub-domains of itself (e.g., sd1.attacker.com, ... , sdF1.attacker.com), causing the resolver to send $2n_1$ additional queries (step 4) to resolve the IPv4 and IPv6 addresses of these delegated name servers. The attacker server then responds to these with $2n_1 = F1$ referral responses (step 5), each with $n_2$ delegations. This results in a total of $2 \cdot F1 \cdot n_2$ delegations, each of which is a name of a fake TLD server (e.g., ns.fake1, ns.fake2, .... , ns.fakeF1xF1x2). $F1$ is bounded by the *max-recursion-queries* parameter of BIND, to 74, and $n_2$ to 135 by the $n_{max}$, resulting in a maximum amplifier fan-out of $74 \cdot 270 = 19,980$ requests. This can potentially lead to a PAF of 3,240, if the target authoritative servers revert to TCP.

This experiment shows a huge discrepancy between the measured and calculated victim cost (81,428 vs. 239,760). This is because the resolver has to send and receive 19,980 requests at the same time, which it fails to do, causing the loss of many request packets. To successfully attack the root (or a TLD), the attacker should combine this technique with the one presented in Fig. 5a, using many different resolvers, all of which direct their requests to the target.

**Long-lived attack simulation.** The discussion so far has focused on the attack power as a result of one client request. Since the attack uses nonexistent domain names, the cache mechanisms do not help, and the attack is long-lived. To show this we simulate a long-lived attack using variant b, which does not interact with external authoritative servers; hence we could test it on our setup without leaking any attack packets outside the virtual lab. As shown in Table 2, $F$ of this variant is 74; thus we include 37 name servers names in each NS referral response. We monitor the packets processed in both the recursive resolver and the victim authoritative server in the test-bed. We used the *resperf* tool [30] on the client machine (acts as the attacker) to send a query stream consisting of many unique DNS 'A' requests to sd*.attacker.com. As shown in Figure 7 (see the 'Original BIND' line), 10,000 attacker requests result in 1,500,319 packets involved in the recursive resolutions, producing a constant PAF of 75x. Here each client query ends with a SERVFAIL, but the recursive resolver's cache is filled with 740,000 NXDOMAIN records (each client request triggers 37 IPv4 resolutions and 37 IPv6 resolutions), and 10,000 NS records. Thus both have a large PAF, which causes very quick saturation of the cache and the memory.

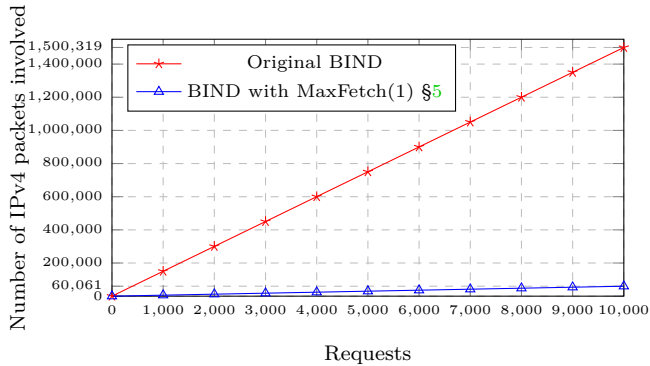

Figure 7: Simulation of long-lived NXNSAttack against an SLD authoritative server. The original BIND implementation has a constant PAF of 75x as compared to 3x in MaxFetch(1). (see §5). Recall that the attacker cost is $2 \cdot \#\text{requests}$.

### 4.4 Public DNS servers

Here we measured the amplification factor attained when each of several public DNS resolvers (such as Cloudflare, Google, Quad9, etc.) is used as the resolver in the amplifier. The DNS software in the public resolvers varies, and some have their own proprietary implementation. We used attack variation b, in which an SLD authoritative is the victim, and we used our own ns.victim.com as the victim. Since we cannot really mount an attack using a public DNS server, we tested each one with one client request at a time, for several requests, until the maximum firepower was found. The results are given in Table 3. We could not test variants a and c because these require monitoring the recursive resolver or the TLD/ROOT servers. To this end, we deployed a 'malicious' name server that responds to queries for `xxx.attacker.live` and sent a few queries to each of the public resolvers, requesting `sdX.attacker.live`. For each such request, our name server, ns.attacker.live, responded with a referral response with a different number of victims and victim sub-domains (all residing in our name server, `victim.online`). For each request sent, we monitored how many requests arrive at the victim name server. All the public DNS resolvers that we tested exhibited a large PAF on a single request of the NXNSAttack. Some have a higher PAF than the one observed in BIND for this variant (74x).

| Public DNS recursive resolver (IP) | Max # of delegations = F/2 | Victim cost $C_v^{pkt}$ | PAF |
|---|---|---|---|
| CloudFlare (1.1.1.1) | 24 | 96 | 48x |
| Comodo Secure (8.26.56.26) | 140 | 870 | 435x |
| DNS.Watch (84.200.69.80) | 135 | 972 | 486x |
| Dyn (216.146.35.35) | 50 | 408 | 204x |
| FreeDNS (37.235.1.174) | 50 | 100 | 50x |
| Google (8.8.8.8) | 15 | 60 | 30x |
| Hurricane (74.82.42.42) | 50 | 98 | 49x |
| Level3 (209.244.0.3) | 135 | 546 | 273x |
| Norton ConnectSafe (199.85.126.10) | 140 | 1138 | 569x |
| OpenDNS (208.67.222.222) | 50 | 64 | 32x |
| Quad9 (9.9.9.9) | 100 | 830 | 415x |
| SafeDNS (195.46.39.39) | 135 | 548 | 274x |
| Ultra (156.154.71.1) | 100 | 810 | 405x |
| Verisign (64.6.64.6) | 50 | 404 | 202x |

Table 3: Firepower and PAF of public resolvers as a response to a single request in the NXNSAttack.

### 4.5 NXNSAttack vs. NXDomain Attack and its effects on the DNS system

Both NXDomain and the NXNSAttack use non-existing domain names to bypass the recursive caches and reach different name servers. While the NXDomain attack (*water torture* [3, 20, 33]) is easier to launch because it does not require a malicious authoritative server. The NXNSAttack is however, more powerful in two aspects; packets/bytes amplification factor, and amount and type of cache records consumed. Its PAF ranges from 74x to 1602x, in contrast to 3x in the NXDomain attack. The NXNSAttack consumes memory and 'NS', 'NX' and even 'A' (in variant c) cache records, while the NXDomain attack consumes only 'NX' cache records at a much slower pace. Note that some ISPs have disabled negative caching due to the increased pervasiveness of one-time signals and disposable domains [13], thus eliminating the cache growth caused by the NXDomain attack. In conclusion, since large recursive resolvers were knocked down by the NXDomain attacks [29, 31], they are more likely to be knocked down by the NXNSAttack.

**Attack effectiveness comparison.** While variant b of the NXNSAttack is the least effective, with the smallest PAF, and is also likely to have the smallest cache consumption rate, it is the only variant we can easily compare against the NXDomain attack in a stress test in our setup. We used the same setup as in the long-lived test in Section 4.3.

In the comparison we measured MaxQps, the maximum rate of attacker client requests before the victim resolver or the authoritative server starts to lose requests. We prepared a file containing one million requests for each attack (each having different bogus requests to instigate the attack) and used it as input to the *resperf* stress tool by Nominum [30], running on the client. (We did not use BIND *queryperf* [12] because it has been reported [30] to produce poor results.) The MaxQps throughput is determined as the point at which the server starts dropping queries and the response rate stops growing, indicating that the server capacity has been exceeded.

The results show that the MaxQps of the BIND recursive resolver significantly degrades under the NXNSAttack, with a peak of 932 Qps. The resolver throughput under the NXDomain attack is 3708 Qps. This mainly attests to the much higher PAF of the NXNSAttack, which requires much fewer malicious client requests to saturate the resolver. As a reference, the max throughput that we measured under non-attack traffic (e.g., a campus DNS trace and top million domains) varies from 6,000 (in the case that most of the requests are not cached) to more than 100,000 Qps (where most of the requests are already in the cache).

### 4.6 Saturating the DNS server

We do not have access to a real authoritative or real resolver servers to show how they fail under attack. As an alternative, we measured the maximum rate of the NXNSAttack type requests that each such server installed on a strong EC2 machine can handle before losing requests. Since this rate of requests is easily attained by the NXNSAttack, we deduce that the attack can easily take down these servers. We used the same setup as in Sections 4.2, 4.3 and 4.5 except of using a xlarge EC2 machine instead of larege EC2 machine (again 4 vCPU with 16GB memory) with BIND 9.12.3 in both resolver mode and authoritative mode. In resolver mode it starts to lose requests at a rate of 932 client requests per second, as in Section 4.5 (with the same requests that are issued by attacking clients in the NXNSAttack). In this experiment, we observed a large difference in CPU resources utilization between the victim and the attacker: the victim 4 vCPU resolver load exceeded 390%, while at the same time, the attacker's authoritative 1 vCPU load was only 3%. In authoritative mode we fed the authoritative two different streams of requests. The first, a stream of 'A' requests, caused the server to start losing requests when a rate of 68,208 rps was reached. The second, a stream of NXDOMAIN random requests, identical to those sent to an authoritative victim in our attack (e.g., in step 4 in Figure 3), reached a maximum rate of 65,418 rps before beginning to lose requests. Therefore, in our attack 1,000 client requests per second (with PAF=x75) would be sufficient to overwhelm this authoritative name server.

## 5 Attack Mitigation: MaxFetch($k$)

### 5.1 Possible and Existing Measures

Several different approaches may be taken to mitigate and reduce the NXNSAttack effect. While MaxFetch(k) is the simplest to integrate and directly patches the problem, we briefly mention few approaches, before diving into the details of MaxFetch(k) in the following subsections:

**MaxFetch(k):** Do not resolve all the name server domains in a received referral response at once, but rather, $k$ per each original client request. See details below.

**Abnormal behavioral analysis:** In the spirit of IPSs, it is possible to monitor the referral messages incoming to resolvers and identify and detect authoritative name servers that send many malicious NS referral responses. One indicator could be abnormally large referrals for zones that appear only once or infrequently. Heavy hitter and distinct heavy hitter algorithms, such as in [9], may be used to detect zones with many bogus sub-domains and filter only the malicious requests. Note that to evade blocking, malicious name servers can dynamically change their name and IP address (in the same manner as malicious C&C servers do). The disadvantage of this approach is that operators will have to deal with yet one more package and the upgrade path is not clear.

**NX replies detection:** One unique abnormal behavior of our attack is that the resolver (for example in Fig. 3) receives nonexistent ('NX') replies while resolving a name server name which appeared in a referral response. This cannot happen in normal operation unless there is a configuration error. A client request that results in one or more such 'NX' responses may be failed [7].

**Rate and other limiters:** Following the NXDomain attack, recent versions of BIND have new manual rate limiting features designed to throttle queries from a resolver to authoritatives that are under attack. These rate-limiters, (e.g., fetch-limits, fetches-per-server, and fetches-per-zone [11]) are, however, a double-edged sword, and can become a way to DDoS an authoritative by issuing many requests to hit the threshold and then block legitimate requests. Moreover, setting a rate-limit per authoritative zone or per authoritative name server does not protect the recursive resolver from the NXNSAttack.

**DNSSEC:** Using DNSSEC-Validated Cache as suggested in RFC 8198 [8] enables the resolver to iden-

tify malicious bogus domain requests before issuing the request. To accomplish this, DNSSEC "metadata" in the form of NSEC(3) and RRSIG records must be used. NSEC provides the main benefit by pointing out to the resolver ranges of sub-domain names that are nonexistent and therefore is able to drop domain requests that fall in them, without making the query itself [36]. This can be combined with the above NX replies detection method.

**Max Breadth:** The MaxFetch($k$) proposal mitigates and significantly reduces the PAF (and BAF) of the attack; however, the attack still consumes large amounts of memory and cache (NX, NS records) per client request, in particular variant $c$. An additional approach is to adopt recommendations to restrict the breadth, i.e., the number of delegation name servers in an NS record of a domain/zone (all of which are included in a referral response). This restriction is supported by the observations made in §6; in particular, Fig. 14 shows that about one-hundredth of a percent of the top 1M domains have more than 13 name servers in a referral response, and less than one percent have more than 7. The limitation should be a function of the level of the zone and of the authoritative name server from which the referral that creates the NS record arrives. Thus, for an SLD zone, a default restriction of 4 might make sense. Investigating the exact limits and effects of this MaxBreadth proposal is beyond the scope of the current paper.

## 5.2 MaxFetch($k$)

We propose to amortize the resolution of multiple delegations for a zone over multiple requests that use that zone, one or a few ($k$) resolutions per request, rather than resolving all the delegations of the zone at once, when the referral for the zone first arrives. Thus, in general, in the resolution of each client request, while using an already resolved delegated name server, the resolver resolves the IP address of an additional $k$ delegated name servers to be ready for future requests. This process continues until all the delegations provided in a referral response are resolved. Several variations are possible in this scheme, for example; start with $k$ concurrent resolutions of referred name server names upon receiving the first referral response within the resolution of a client request. Then, on each subsequent client request that results in the same referral, make one (or more) additional name server name resolutions.

We modified the BIND 9.12.3 resolver algorithm to implement MaxFetch(1). The max number of external fetches (additional resolutions) we enforce at each level is configurable. MaxFetch($k$) allows the resolution of $k$ additional delegations that do not have an associated IP address, per request. In MaxFetch(1), a resolver that uses a zone $z$ while resolving a request checks whether there are unresolved delegations for $z$ in $z$'s NS record. If such a delegation is found, the resolver initiates its resolution, while continuing in parallel the resolution of the original request, using an already resolved delegate for zone $z$. Note that the first request that uses zone $z$ (which has also received the corresponding referral response) may have to wait for the resolution of the first delegate if all of them came without a glue record in the referral response (or all are *out-of-bailiwick*). In this case the second request that uses zone $z$ will use the same delegate as used by the first one (one may consider resolving two delegations in the first request, something we have not tested).

It is important to note that MaxFetch(1) does not negatively affect the latency of a request resolution (see latency analysis in §5.3 and §5.4), nor does it disturb the RTT estimation algorithms (such as sRTT). Most recursive resolvers perform latency-wise algorithms to decide which server to query next. However, MaxFetch(1) does not disrupt these algorithms because it allows a resolution of an additional name server that may be selected in the next client request, and after enough requests all the delegations are resolved. The resolution of an additional name server does not add to the latency of a response since each request, except the first, uses a previously resolved name server while issuing the additional resolution in parallel.

In the next sub-sections we evaluate and compare the original BIND and MaxFetch(1). We focus on the impact on the latency and the number of packets, per client request, under normal traffic and under attack.

## 5.3 MaxFetch(1) evaluation under attack

In Figure 7 (§4.1) we compare the PAF of the original BIND to that of the MaxFetch(1) variant, during a long-lived simulated NXNSAttack against an SLD victim. The blue line ($-\triangle-$) shows that the MaxFetch(1) enhancement avoids most of the additional resolutions, since it initiates only two additional requests, one IPv4 and one IPv6 per reqest. Instead of 1,500,319 packets exchanged by the original BIND recursive resolver (as a result of 10,000 malicious client requests), MaxFetch(1) exchanges only 60,061 packets (the measured Mac1Fetch PAF is reduced from $75x$ to $3x$).

We also repeated the stress tests as in §4.5 to measure the maximal number of client queries per second that the BIND resolver is able to sustain under the NXNSAttack with and without MaxFetch(1). As seen in Table 4, BIND with MaxFetch(1) is capable of processing many more attack requests, 3,390 vs. 932 under the NXNSAttack (and 3708 orig. BIND under the NXDomain attack §4.5). We also compared the latency of attack requests with and without MaxFetch(1). The latency values are observed at the attacker client, which that generates requests during a simulation of the NXNSAttack against an SLD victim in our testbed. As seen in Table 4, the average, median, and std latency, under attack, are much better with MaxFetch(1) than without.

|                       | Orig Bind 9.12.3 | MaxFetch(1) |
|-----------------------|------------------|-------------|
| Max requests/sec      | 932              | 3390        |
| Avg. Latency (ms)     | 4.31             | 1.32        |
| Median Latency (ms)   | 4                | 1           |
| std Latency           | 4.51             | 1.37        |

Table 4: Comparing BIND resolver performance under the NXNSAttack with and without MaxFetch(1).

## 5.4 MaxFetch(1) in normal operation

Here we evaluate the recursive resolver operation in practice, with and without MaxFetch(1) under normal operation (without an attack). We measure (i) the latency of client queries and (ii) the number of IPv4 packets processed by the resolver in the resolution process. The purpose is twofold: first, we wish to verify that the MaxFetch(1) modification does not incur query delays or resolution failures (i.e., the number of SERVFAIL and NOERROR responses is not higher than that observed in the original BIND). Second, we wish to measure the impact of the *Out-of-Bailiwick* overhead on the recursive resolver under normal operation, to determine whether the cache mitigates this overhead over time.

### 5.4.1 Datasets

Two datasets are used to study the normal operation of a BIND resolver:

**Dataset $\mathcal{A}$:** A list of the **top million domains** [21]. Here we executed DNS 'A' requests (IPv4 resolution) for the first 100,000 domains in this list.

**Dataset $\mathcal{B}$: Campus DNS trace**. A 24-hour trace of live DNS traffic observed on a campus DNS server. Out of the 1,027,359 queries to domains that do not reside within the campus zone, we took the 386,736 'A' queries, with 10,092 unique ones.

**Ethical Consideration:** Dataset $\mathcal{B}$ is a sequence of DNS queries with their timestamps but without the IP addresses that originated them.

With each dataset, we send its query stream (100,000 queries in Dataset $\mathcal{A}$, and 386,736 queries in Dataset $\mathcal{B}$) to both original BIND and BIND with MaxFetch(1). The 1GB resolvers' cache is empty at the beginning of each experiment, and it can store all the responses received in each experiment. We record the traffic between the recursive resolver and the authoritative hierarchy, as well as collecting the BIND statistics.

|                       | Dataset $\mathcal{A}$ (100K top domains) | | Dataset $\mathcal{B}$ (Campus trace) | |
|-----------------------|------------------|-------------|------------------|-------------|
| Resolver Impl.        | Original BIND    | Max-Fetch(1) | Original BIND    | Max-Fetch(1) |
| Total Req.            | 100,000          | 100,000     | 386,691          | 386,691     |
| Unique Req.           | 100,000          | 100,000     | 10,092           | 10,092      |
| Total recursion packets | **747,494**    | **650,864** | **454,032**      | **422,946** |
| NOERROR               |                  |             | 363028           | 363031      |
| SERVFAIL              |                  |             | 18911            | 18910       |
| NXDOMAIN              |                  |             | 4752             | 4750        |
| Latency (ms)          |                  |             |                  |             |
| Mean                  | **157.37**       | **155.95**  | **41.50**        | **40.97**   |
| Median                | 53               | 52          | 13               | 13          |
| Std                   | 298.63           | 293.37      | 101.03           | 95.81       |

Table 5: Comparing original BIND and BIND with MaxFetch(1) during the resolution of the query streams in Datasets $\mathcal{A}$ and $\mathcal{B}$.

### 5.4.2 Results

**Resolution overhead.** We start by measuring the drop in resolution cost introduced by Max1Fetcch in normal operation (see §2.2). Figure 8 and the fourth row (Total recursion packets) in Table 5 show the number of packets processed by the recursive resolver (with and without MaxFetch(1)) in each of the datasets. Using original BIND, the resolver exchanges 14.84% more packets in the resolution of the queries in Dataset $\mathcal{A}$ than it does using the MaxFetch(1) variant (747,494 vs. 650,864). Similarly, for Dataset $\mathcal{B}$ (campus DNS trace), original BIND exchanges 7.34% more packets (454,032 vs. 422,946).

As seen by the green overhead lines ($-\diamond-$) in Figures 8a and 8b, MaxFetch(1) saves more than 50% of the resolution cost in the first 1000 requests, when the cache has not yet filled up. The lines show the resolution drop in percentages, $((\frac{Packets_{orig}}{Packets_{MaxFetch(1)}} - 1) \cdot 100)$. The gap decreases as more requests are processed and the cache is filled up with name server resolutions that are shared by multiple requests. After 20K requests, the gap remains stable at around 15% for Dataset $\mathcal{A}$, and 7% for Dataset $\mathcal{B}$. Furthermore, MaxFetch(1) does not result with more SERVFAIL, NOERROR, or NXDOMAIN than original BIND in the resolution of the 386,691 queries in Dataset $\mathcal{B}$ (Fifth row in Table 5).

**Latency.** The last row in Table 5 shows the average, median and std latency, in both data sets, with and without MaxFetch(1). The response time is slightly faster using MaxFetch(1): 157.37ms using original BIND vs. 155.95ms using MaxFetch(1) in Dataset $\mathcal{A}$ (top domains), and 41.5ms vs. 40.97ms in Dataset $\mathcal{B}$ (campus trace). Note that in Dataset $\mathcal{B}$, most queries are served by the resolver cache because not all the requests are unique. Thus, when calculating the average, median and std calculations, we exclude queries with zero latency (we consider only 114,570 out of 386,691 queries).

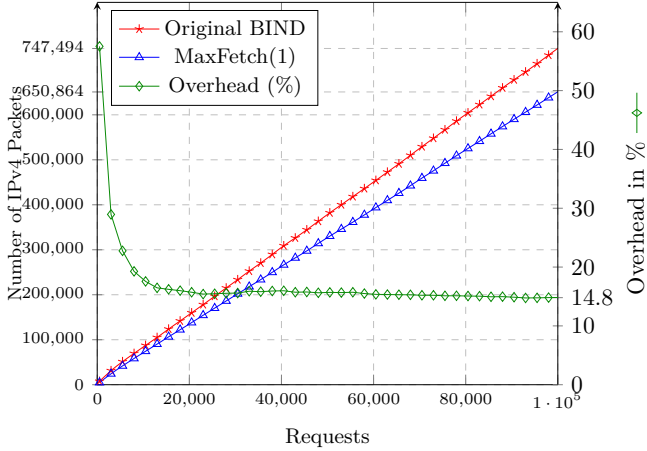

(a) 100K Top Domains

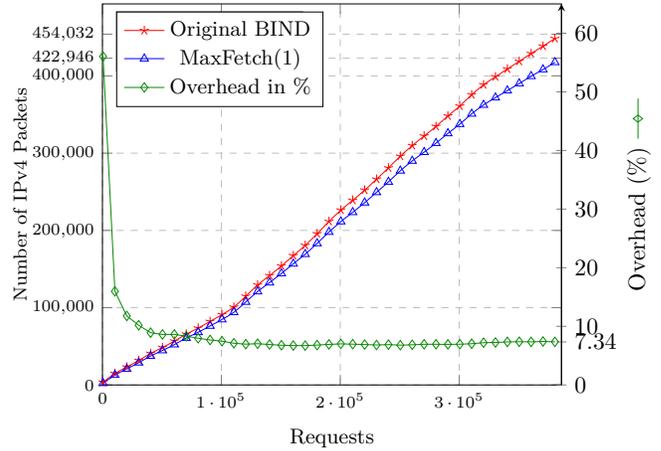

(b) Campus 1-Day Trace

Figure 8: The number of recursion packets exchanged by a BIND resolver (with and without MaxFetch(1)) in the resolution of Dataset $\mathcal{A}$ and $\mathcal{B}$ query streams. The green line ($-\diamond-$) shows the overhead that is relative to MaxFetch(1).

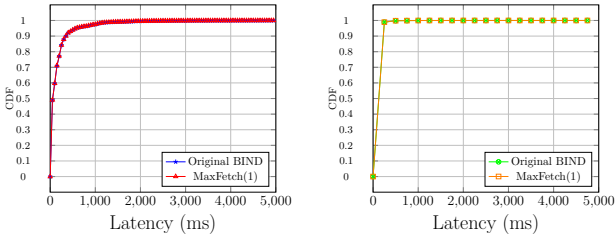

(a) 100K top domains    (b) Campus trace

Figure 9: Latency of queries in Datasets $\mathcal{A}$ and $\mathcal{B}$: Comparison between original BIND and MaxFetch(1).

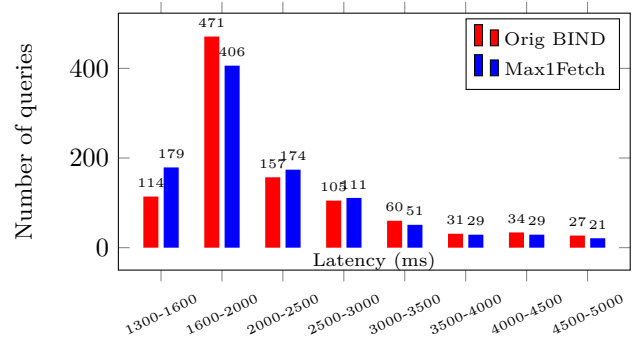

Figure 10: Dataset $\mathcal{A}$, $99^{th}$ percentile BIND latency with and without MaxFetch(1).

Figures 9a shows the cumulative distribution of the queries' latency values for Dataset $\mathcal{A}$, with and without MaxFetch(1), and Figure 9b shows the same for Dataset $\mathcal{B}$. The latency values for both datasets with and without MaxFetch(1) are between 0 and 5 seconds. In both datasets, the original BIND and MaxFetch(1), the CDF lines overlap, exhibiting a nearly identical distribution.

The $99^{th}$ percentile latency distribution in Dataset $\mathcal{A}$ (top domains) is provided in Figure 10. The quantile values (cut points of the $99^{th}$ percentile) for original BIND and MaxFetch(1) are 1,414ms and 1,382ms respectively. Similarly, Figure 11 shows the $99^{th}$ percentile distribution for Data-set $\mathcal{B}$.

Figure 12 presents the latency differences per domain request (between original BIND and MaxFetch(1)) in the top domains dataset. Here, for each request for domain $d$ we calculate $L_{orig}^d - L_{m1f}^d$, where $L_{orig}^d$ is the latency of the query for $d$ using original BIND, and $L_{m1f}^d$ when using MaxFetch(1). Figure 12 shows the distribution of the calculated values (vary from -5000 to 5000), where positive values represent domain requests for which MaxFetch(1) performed faster.

## 6 The Pervasiveness of Out-of-Bailiwick Nameservers

Here we measure the prevalence of domains with *out-of-bailiwick* name servers. We show that the majority of the domains out of the top 1M popular sites [21] have *out-of-bailiwick* name servers. We performed two controlled experiments to monitor the resolvers' operation and to examine the NS referral responses in the resolutions of these domains.

In the first controlled experiment we measured how many recursive resolutions a BIND based resolver performed when resolving each of the top 1M domains. The cache was cleared before issuing each client request. In an attempt to capture the difference between the number of resolutions performed per client request in practice vs. theory (as in Section 2 Figure 1 versus Figure 2). That is, at each level of the hierarchy one resolution is not counted. Figure 13 shows the cumulative distribution of domains that trigger additional resolutions (fetches). The figure shows that 60.22% of the domain requests initiate more than one additional fetches. We see that 374,498 domain requests do not initiate any additional resolutions (38.34%, note that we count only requests with NOERROR responses).

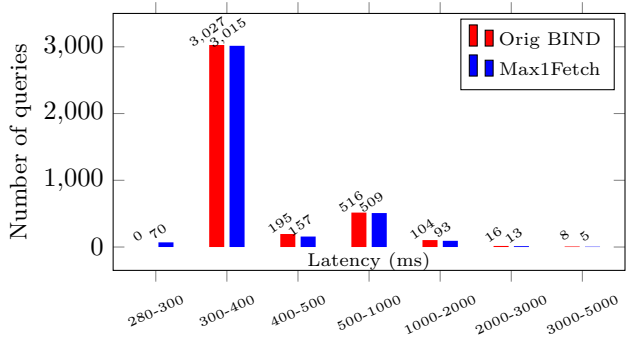

Figure 11: Dataset $\mathcal{B}$, $99^{th}$ percentile BIND latency with and without MaxFetch(1).

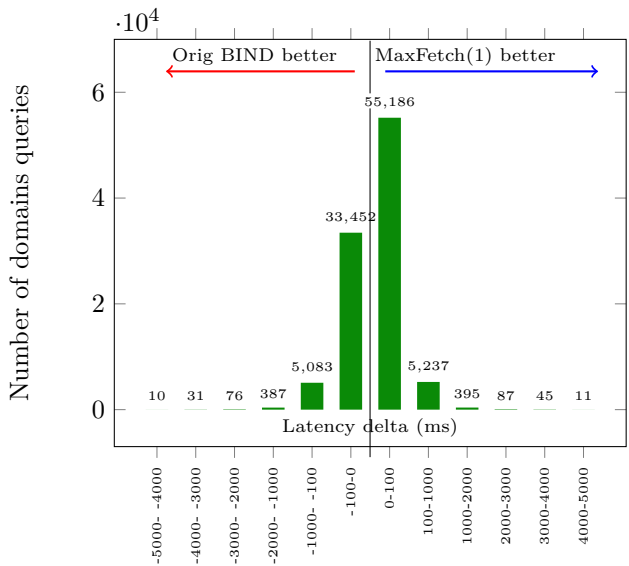

Figure 12: 100K websites dataset: OrigBIND – MaxFetch(1) latency per domain histogram.

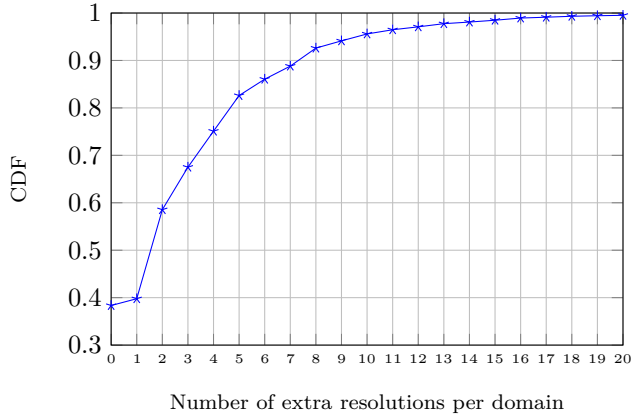

Figure 13: Number of domains that their resolution incurs initiation of extra resolutions by BIND (over top million domains).

In the second experiment, we recorded the communication between the recursive resolver and the authoritative structure during the resolution of the 1M domains. In this case, we did not focus on the BIND operation, but rather inspected the NS referral responses received from the authoritative hierarchy in order to measure: (i) how many name servers are returned for each domain, (ii) how many name servers are not provided with their corresponding IP addresses (missing glue records), and (iii) which name servers are *out-of-bailiwick*. When counting the number of *out-of-bailiwick* name servers, we consider both definitions as we discuss in §2.2 (RFC 8499). The first, strict definition describes a name server whose name is subordinate to the owner name of the NS resource record (e.g., ns.child.example.com as name server for the domain 'example.com'). The second, wider definition refers to a name server's name that is subordinate to the zone origin and not subordinate to the owner of the NS resource record (e.g., ns.another.com as name server for the domain 'example.com').

We start by counting the number of name servers for each domain; see Figure 14. While most of the domains have two name servers, 33% have three or more. Results show that the top million domains have an average of 2.52 name servers per domain.

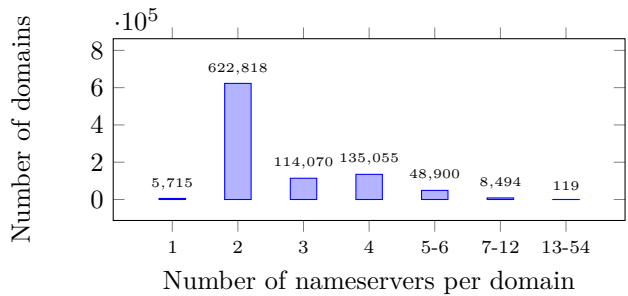

Figure 14: Number of duplicates of each name server per domain over top million domains.

| Measurement | Number |
|---|---|
| Requests | 1,000,000 |
| Answers | 1,000,000 |
|   NXDOMAIN | 20,025 |
|   SERVFAIL | 20,110 |
|   NOERROR | 959,865 |
|     CNAME Response | 1,717 |
|     Empty Response | 11,498 |
|     Domains with nameservers (valid) | 946,650 |
|       Domains that all their NSs with glue (IP) | 342,429 |
|       Domains that all their NSs w/o glue. | 567,450 |
|     Total name servers in answers | **2,394,475** |
|       In-bailiwick name servers (strict def.) | 70,596 |
|       Out-of-bailiwick name servers (strict def.) | 2,323,879 |
|       In-bailiwick name servers (wider def.) | 1,081,876 |
|       Out-of-bailiwick name servers (wider def.) | **1,312,599** |
|       name servers with glue records | 869,140 |
|       name servers w/o glue records | **1,525,335** |

Table 6: Pervasiveness of authoritative name servers with missing glue records over the top million domains.

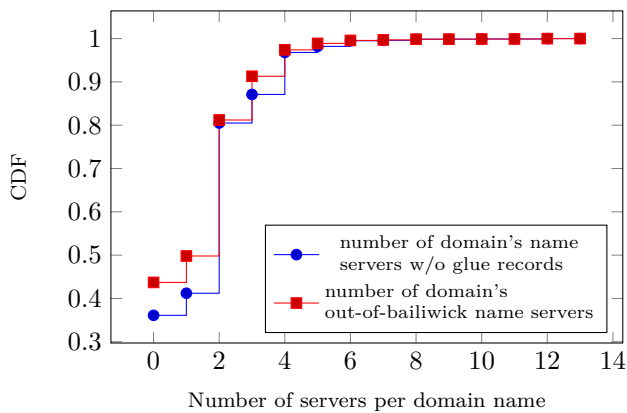

Figure 15: CDF of out-of-bailiwick name servers per domain.

We show the results in Table 6. Only 869,140 out of 2,394,475 (36.3%) name servers that appear in the NS referral responses of the 1M domains are both in-bailiwick and include a corresponding IP address (glue record). 1,525,335 (63.7%) name servers are missing a corresponding glue record, of which 1,312,599 are out-of-bailiwick, showing that some *in-bailiwick* name servers are not provided with their glue records by their parent authoritative name servers. Here we refer to the wider definition of *bailiwick*; the results show that most authoritative name servers provide glue records according to this definition. Additionally, according to the strict definition of *in-bailiwick*, we found that only 70,596 name server out of 2,394,475 (2.95%) are *in-bailiwick*, i.e., their name servers names are within the domain name (for example, 'ns.example.com' as a name server for the domain 'example.com').

The blue line (−•−) in Figure 15 shows the distribution of the number of name servers without a glue record per domain. For the majority of the domains (567,450 out of 946,650 domains with NOERROR responses, 59.94%), *all* their name servers are received without a corresponding glue record (in the NS referral response from the TLD, or sometimes from an SLD). One reason for this high number of domains with *out-of-bailiwick* name servers is that many domains outsource their DNS authoritative service to the same vendors. Out of the 1 million we tested, 218,747 (21%) domains use `ns.cloudflare.com` and 129,789 use `domaincontrol.com`.

## 7 Related Work

Luo et al. [20] analyze the prevalence and characteristics of the NXDomain and water torture attacks. Using one month of real-world DNS traffic, they compare the attack behavior with DGA malware and disposable services.

Recently the DNS infrastructure is facing abuse by various entities which use it for applications for which it was not intended. In this case, a large volume of temporary domain names (aka disposable domains [13]) is commonly used to help these services to communicate via DNS queries. A study [13] from large scale DNS traffic shows that 60% of all distinct resource records observed daily are disposable. Hao et al. [5] examine the negative impact of disposable domains on recursive caching. They propose a classification based on domain name features to increase the cache hit-rate.

Maury [22] presents a different attack that also exploits the delegations of name servers in a referral response. However, the attack (called the iDNS attack) PAF is at most 10x. In iDNS the attacker's name server sends self-delegations (back and forth to the attacker's name server) up to an infinite depth. A major difference from our work is that the glueless name servers in the iDNS attack are never used against an external server such as a victim name server. Some measures have been taken by different DNS vendors such as BIND and UNBOUND following the disclosure of iDNS described in [22]; however these measures do not affect and do not weaken the NXNSAttack.

Wang [35] focuses on the DNS security implications of glue records. He describes how recursive resolver implementations such as BIND and Unbound treat glue records, but the focus is on cache poisoning vulnerabilities rather than the impact on the recursive performance, which is the focus of the current paper.

Muller et al. [28] perform a comprehensive measurement using the RIPE atlas to analyze how recursive resolvers select which name server to interact with, out of a set of multiple authoritative servers. The focus is on how and when the recursive resolvers *query* a set of multiple authoritative servers, while in this paper we extend the discussion and focus on how and when recursive servers *resolve* the IP addresses of a set of authoritative name servers. In another work [27], Moura et al. analyze the root DNS service during a specific DDoS attack. However, the analysis refers to authoritative servers rather than recursive behavior. In a recent work [26], Moura et al. measure and show the impact of the caching and long TTL on dissecting DNS defenses during a DDoS attack.

## 8 Disclosure

After discovering the NXNSAttack, we initiated a responsible disclosure procedure. The following vendors and DNS service providers were approached and have patched their software and servers, most of them using the MaxFetch($k$) approach: ISC BIND (CVE-2020-8616), NLnet labs Unbound (CVE-2020-12662), PowerDNS (CVE-2020-10995), CZ.NIC Knot Resolver (CVE-2020-12667), Cloudflare, Google, Amazon, Microsoft, Oracle (DYN), Verisign, IBM Quad9 and ICANN. Akamai among others, seems to have been patched or non-vulnerable to a one variant of the attack which we checked. Here is a quote from one of the large parties in the disclosure: "Due to this attack's potential to incapacitate a target's authoritative name server with very little effort on behalf of the attacker, we've rated the original report a High severity".

## 9 Conclusions

You never know what you might find when you go searching for your lost donkey. We started off researching the efficiency of recursive resolvers and their behavior under different attacks, but we ended up finding a new, devastating vulnerability, the NXNSAttack.

The key ingredients of the new attack are (i) the ease with which one can own or control an authoritative name server, (ii) the use of nonexistent domain names for name servers, and (iii) the extra redundancy placed in the DNS structure to achieve fault tolerance and fast response time.

We note that some of the possible remedies, such as various rate limiters, are a double-edged sword; a sophisticated attacker may use them to deny service to legitimate clients, by hitting the limiter's thresholds with malicious requests.

Notice that DoH (DNS over Http) is irrelevant to this paper because it deals with the communication channel between a client and its recursive resolver while we focus here on the communications between the recursive resolver and the authoritative structure.

**Acknowledgements:** We would like to thank the referees for very helpful comments and feedback, and Michael McNally, and Cathy Almond of ISC, Ralph Dolmans, Wouter Wijngaards and Benno Overeinder of NLnet Labs, and Petr Špaček of NIC.CZ for their help and cooperation in the disclosure procedure, as well as Eyal Ronen and Yair Kaldor for their help in this project.